# Broadband Linear-Dichroic Photodetector in a Black Phosphorus Vertical p-n Junction

Hongtao Yuan,[1,2] Xiaoge Liu,[1] Farzaneh Afshinmanesh,[1] Wei Li,[1] Gang Xu,[1] Jie Sun,[1] Biao Lian,[1] Alberto G. Curto,[1] Guojun Ye,[3] Yasuyuki Hikita,[1,2] Zhixun Shen,[1,2] Shou-Cheng Zhang,[1,2] Xianhui Chen,[3] Mark Brongersma,[1,2] Harold Y. Hwang,[1,2*] Yi Cui[1,2*]

[1]Geballe Laboratory for Advanced Materials, Stanford University, Stanford, California 94305, USA

[2]Stanford Institute for Materials and Energy Sciences, SLAC National Accelerator Laboratory, Menlo Park, California 94025, USA

[3]Hefei National Laboratory for Physical Science at Microscale and Department of Physics, University of Science and Technology of China, Hefei, Anhui 230026, China

**The ability to detect light over a broad spectral range is central for practical optoelectronic applications, and has been successfully demonstrated with photodetectors of two-dimensional layered crystals such as graphene and MoS$_2$. However, polarization sensitivity within such a photodetector remains elusive. Here we demonstrate a linear-dichroic broadband photodetector with layered black phosphorus transistors, using the strong intrinsic linear dichroism arising from the in-plane optical anisotropy with respect to the atom-buckled direction, which is polarization sensitive over a broad bandwidth from about 400 nm to 3750 nm. Especially, a perpendicular built-in electric field induced by gating in the transistor geometry can spatially separate the photo-generated electrons and holes in the channel, effectively reducing their recombination rate, and thus enhancing the performance for linear dichroism photodetection. This provides practical functionality using anisotropic layered black phosphorus, thereby enabling novel optical and optoelectronic device applications.**

[*] Corresponding author: hyhwang@stanford.edu, yicui@stanford.edu.

Confined electronic systems in layered two-dimensional (2D) crystals are host to many emerging electronic, spintronic and photonic phenomena,[1, 2, 3] including quantum Hall and Dirac electrons in graphene[4, 5, 6] and topological surface states in topological insulators[7, 8]. Experimentally identifying new functionalities of two-dimensional materials is a challenging and rewarding frontier, enabled by recent advances in materials and device fabrication. One example is the valley polarization control using circularly polarized light in the non-centrosymmetric $MoS_2$ monolayer and resulting potential valleytronics applications.[9, 10, 11] Other examples include recent demonstrations of novel electronic and optoelectronic applications of the well-known layered material black phosphorus (BP), such as high-mobility field effect transistors and linear-polarization dependent optical absorption.[12,13,14] Therefore, further discovering new properties and functionalities utilizing known layered materials is of practical importance and great current interest.[14, 15, 16, 17, 18, 19, 20, 21, 22, 23, 24, 25, 26]

As a potential functionality for layered 2D materials, linear dichroism (LD) is an electromagnetic spectroscopy probing different absorption of light polarized parallel or perpendicular to an orientation axis. It directly depends on the conformation and orientation of material/device structures, where they are either intrinsically oriented in an anisotropic crystal structure[27, 28] or extrinsically oriented in anisotropic device patterns[29, 30]. Compared to the hexagonal in-plane lattice in other 2D materials such as graphene and $MoS_2$, which are insensitive to the linear polarization of incident light, the layered BP crystal with a rectangular in-plane lattice has a highly-anisotropic structure along the $x$ and $y$ directions (defined in Fig. 1a), where every two rows of P atoms alternately puckers up and down to form an "armchair"-like geometry only along the $x$ direction. Therefore the electrons and photons in BP can behave in a highly anisotropic manner within the layer plane.[18, 19] In contrast to existing photodetectors for linear dichroism detection based on extrinsic geometric effects such as with wire-grid patterns,[30, 31, 32, 33] BP with a mirror reflection symmetry only in the $y$ direction (Supplementary Fig. S1a) offers an exciting opportunity to incorporate intrinsic crystal anisotropy for linear-polarization-sensitive photodetection. In this Article, we demonstrate a linear dichroic photodetector based on BP transistors, where the polarized light is absorbed differently along two in-plane crystal axes across a very broad spectral range from about 400 nm to 3750 nm. The optical selection rules in BP allow the broad-band absorption of the light polarized preferential in the $x$ direction, and leads to a new degree of freedom to control/manipulate electronic and optoelectronic properties in layered BP. In particular, utilizing the vertical p-n junction induced by an ionic gel gated transistor, we spatially separate the photo-generated electron-hole pairs in the channel, reducing their recombination rate and thus enhancing the photoresponsivity of the photodetector.

Similar to the layered structure of other 2D materials, BP is composed of stacking the monolayer structure (four-formula $P_4$ block unit) along the $z$-axis (Fig. 1a). Different from the layered crystals with $sp^2$ bonding and a flat in-plane lattice, the BP monolayer, puckered along the $x$ direction due to the $sp^3$ hybridization, creates an anisotropic band structure. We performed angle resolved photoemission spectroscopy (ARPES) and theoretical band calculations to study the structure of the band dispersion anisotropy of BP. In the experimental band structure obtained by ARPES (Figs. 1c and 1d), the valence band maximum of bulk BP resides at the Z point, the center of the $k_z = \pi$ plane in the Brillouin zone (Fig. 1b).[34] Furthermore, the bands along Z-L ($k_x$) and Z-P ($k_y$) directions have different evolving slopes. The elliptic pockets in the constant energy contours (Fig. 1d) directly reflect the anisotropic effective mass and free carrier velocity along the two perpendicular in-plane directions. Band calculations clearly indicate a direct band gap of 0.33 eV for bulk BP. The direction with small effective masses (0.12 $m_0$ for holes and 0.11 $m_0$ for electrons) is the $k_x$ direction (the puckering $x$ direction in real-space). In contrast, the valence (conduction) band along the $k_y$ direction is a relatively flat band and gives a larger hole (electron) effective mass of about 0.81 $m_0$ (1.3 $m_0$), reflecting a large electronic anisotropy along Z-P and Z-L directions (Fig. 1c).[12] Considering the orbital component within this $sp^3$ hybridization (partial density of states shown in Fig. 1e), the band dispersions near the valence band maximum and the resulting band gap mainly originate from the $p_z$ orbital localized along the $z$ direction.

A symmetry analysis from the band structure for the optical selection rules can explain why and how BP exhibits linear dichroism and anisotropic absorption along the two perpendicular planar directions. As shown in Supplementary Fig. S1a, the BP crystal structure possesses inversion symmetry (parity) and mirror reflection symmetry ($M_y$) only in the $y$ direction. Electron states at high symmetry points (*e.g.*, Z, Γ) in the Brillouin Zone can thus be labeled by these two symmetries, and correspondingly there are two optical selection rules: 1) Parity selection rule: Since the photons carry parity $-1$, the first optical selection rule is that the optical absorption by an electron can only occur between electron states (bands) with opposite parities; 2) Polarization selection rule: This rule concerns the polarization of the normally incident light: Under mirror reflection ($M_y$), the field of polarized light ($E_y$) in *y*-direction changes sign ($E_y \rightarrow -E_y, B_x \rightarrow -B_x$), while that of polarized light ($E_x$) in *x*-direction is unchanged ($E_x \rightarrow E_x, B_y \rightarrow B_y$). Therefore, the absorption of the *x*-direction (*y*-direction) polarized light is only allowed between electron states with the same (opposite) eigenvalue of $M_y$ (for example, +1 or -1). We note that the second rule applies also for all electron states in the *y*-direction momentum $k_y = 0$ plane.

According to the optical selection rules, we find the lowest energy optical transition in BP occurs at the Z point across the gap (~0.33 eV) between the valence band Vb1 and the conduction band Cb1 with opposite parities (Fig. 2a).[27] Both bands have $M_y = +1$ at Z point and in the $k_y = 0$ plane. The electron states in the $k_y = 0$ plane (Z-L direction in Fig. 2a) are therefore forbidden to absorb light polarized in the *y*-direction. In contrast, states in the $k_x = 0$ plane (Z-P direction in Fig. 2a) in general have no $M_y$ symmetry, and thus the absorption of both *x*- and *y*-direction polarized light is symmetry allowed. Overall, the absorption of polarized light in *y*-direction is forbidden at 0.33 eV and also significantly reduced at higher energies, which leads to linear polarization dependent absorption spectrum in BP. Note that the pair of bands Vb1↔Cb1 can contribute to the intrinsic light absorption in a broad range of photon energy from 0.33 eV to 4 eV, and thus supports a wide continuous polarization dependent absorption spectrum. However, optical absorption channels between other bands will be involved for photon energies above 1.2 eV, where the polarization-sensitivity is either weakened or strengthened. Based on the band structure calculations (Fig. 2a), we find that all the photon absorption channels up to around 3.0 eV favor the absorption of *x*-polarized light, while the band pairs of Vb3-Cb1 and Vb1-Cb4 favoring *y*-polarized light are involved in absorption processes above 3.0 eV (see Fig. 2a and Supplementary Section 3). Therefore, in principle the polarization dependence of the absorption continuously covers a wide range from 0.33 eV up to ~3.0 eV (3750 nm – 400 nm in wavelength).

To experimentally verify the linear dichroism of BP, we performed polarization dependent absorption and reflection measurements on cleaved BP flakes (details are given in Supplementary Section 3). The crystal orientation of mechanically-cleaved BP samples is investigated by systematic studies using high-resolution transmission electron microscopy (Supplementary Section 1). In the infrared (IR) regime (Fig. 2b), the reflection spectrum with the incident light polarized along the *x*-axis of the BP crystal has a sudden drop with photon energies larger than the band gap (0.33 eV) of BP. In contrast, the reflection of light polarized along the *y*-axis shows no decrease near the band edge. One can clearly see the gradual evolution of the reflection spectra (from 30% to ~90%) using light with changing polarization angles from *x* to *y* directions. This implies that more photons are absorbed with incident light polarized along the *x*-axis relative to the case with light polarized along the *y* axis, consistent with our theoretical analysis and recent optical studies.[12,18,19] Similar polarization dependent phenomena are observed for visible light (Fig. 2c); the transmittance with the light polarized along the *x*-axis is low (42% at 2.3 eV) and the flake appears darker with more absorption (inset of Fig. 2c). By contrast, when light is polarized along the *y*-axis, the transmittance at 2.3 eV is as high as 80%, and the flake appears transparent. A video in the Supplementary Information clearly indicates the gradual changes of the transmittance during continuous rotation of the polarization axis. Such a linear dichroism is consistent with recent reports on the linear polarization dependent absorption in BP flakes.[12,18,19]

The polarization dependent anisotropic absorption in BP can be further reflected by its photoconductivity. Scanning photocurrent microscopy was performed using a supercontinuum Fianium laser (400-1700 nm in wavelength) to measure the spatially dependent photoresponse of BP devices (first on a Hall-bar patterned BP device as shown in Supplementary Section 4). A small DC voltage ($V_{SD}$ = 0.1 V) is applied between the source-drain electrodes (Supplementary Fig. S4b). As shown in the large-area photocurrent scanning image for the whole channel (Supplementary Fig. S4c), one can see the strong signal of the photocurrent |$I_{ph}$| at two different BP/metal electrode edges and zero photoresponse in the center area of the flake far from electrodes. Supplementary Fig. S4d shows the evolution of photocurrent mapping images (|$I_{ph}$|) at a constant wavelength of 1500 nm as a function of the polarization angle of the incident light, with the laser spot scanning near the metal electrode edges (indicated by the yellow square in Supplementary Fig. S4c). The photocurrent is maximum by showing a strip shape when the incident light is polarized along the *x* crystal axis (defined as 0 ° polarization) and is minimum when the incident light is polarized along the *y* crystal axis (defined as 90 ° polarization), directly indicating the polarization dependent absorption and the resulting linear dichroic photocurrent generation. Note that the photocurrent at two different BP/metal electrode edges has opposite flowing directions (more details are given in Supplementary Fig. S8) and there is a zero photocurrent crossover along the channel. Such a current flow profile and the zero-photocurrent crossing along the current channel can be ascribed to the photo thermoelectric effect under pulsed excitation heating rather than the photovoltaic effects originating from the built-in electric field between the source/drain electrodes. Our systematic measurements indicate that the photo thermoelectric effect dominates the photocurrent generation at zero and low DC voltages ($|V_{SD}| < 0.15$ V) while the photo-voltaic effect starts to dominate the photocurrent process at higher $|V_{SD}|$. More details are given in Supplementary Section 6.

To exclude the possibility that the observed two-fold polarization dependent photocurrent originates from a geometric edge effect at the metal/BP edge, we designed a "ring"-shaped metal electrode as the photocurrent collector (Fig. 3a), in which the photogenerated hot carriers can be collected isotropically where the influence from the orientation of the electrode edge is the same for all polarizations. A small DC voltage ($V_{SD} = 0.1$ V) is applied between the source-drain electrodes so the device works in the photo thermoelectric regime, aiming to see the intrinsic behavior from the anisotropic BP flakes without the photocurrent driven by an external electric field in the photovoltaic regime. Figure 3b shows the full spatial mapping of the photocurrent around the ring electrode at a wavelength of 1500 nm. Focusing the excitation in the round area within the inner diameter of the "ring" electrode, the photocurrent with 0 ° light polarization (*x* crystal axis) is much larger than that for 90 ° light polarization (Fig. 3c), suggesting that the intrinsic polarization dependent photoresponse originates from BP itself. The photoresponsivity (photocurrent normalized to the incident laser power) at 1200 nm is as large as 0.35 mA/W and also has a large contrast ratio (3.5) between the photoresponsivity along the two perpendicular polarizations.

These observations clearly indicate that the incident light in different polarization states travelling through the dichroic BP experiences a varying absorption, directly reflecting the intrinsically anisotropic conformation and orientation of the crystal structure. Related to the total intrinsic vertical optical transition for a single pair of bands (Vb1 ↔ Cb1 transition mentioned above for Fig. 2a), the photoresponsivity to polarized laser excitation with varying wavelength range from 400 nm to 1700 nm directly indicates the wide bandwidth of the linear dichroism detection (Fig. 3c). Compared to reported linear dichroism applications based on a wire-grid polarizer,[30,31] which have largely relied on advances in nanofabrication, our observations on the linear dichroism photodetection demonstrate that layered BP can be used as a potential intrinsic linear dichroism media with broad-band response for practical integrated optical applications. Also interestingly, by combining the highly efficient polarization dependent optical absorbing characteristics with the good transport characteristics, BP could be a promising material for high performance optoelectronic devices. Its relatively small carrier effective masses along the $x$ and $z$ direction give a carrier mobility an order of magnitude larger than typical perovskite absorbers[35, 36] and layered chalcogenides[16, 37]. Consequently, the large carrier mobility directly affects the photoresponse speed of the BP photodetector devices. As indicated in Supplementary Section 7, the photoresponse rise time of the linear dichroic BP photodetector can be faster than 40 micro-seconds (μs), the detectable time-resolution limitation of the preamplifier in our photocurrent measurement setup. This value of 40 μs is clearly faster than the photoresponse time of previously reported photodetector based on layered chalcogenides (normally in the order of milliseconds or even slower)[16, 37].

To enhance the performance of the BP photodetector, we formed the BP flakes into an ionic gel gated electric-double-layer transistor (EDLT), which has been known to be a powerful tool to tune interfacial band bending (perpendicular electric field) and also the Fermi level of channel materials over a large range.[38, 39, 40] Figures 4a and 4b show the ambipolar transfer characteristics, the source-drain current $I_{DS}$ and sheet resistance $R_s$ as functions of gate voltage ($V_G$) for the BP-based EDLT. As a normally-ON transistor with a p-type channel, negative $V_G$ accumulates holes at the gel/BP interface with upward band bending. Positive $V_G$ first depletes the holes away from the interface and with further increase of $V_G$, an electron inversion layer can be induced (Supplementary Section 8). Hall effect measurements (Figs. 4c and 4d) clearly indicate the ambipolar operation in the channel, with the transition from hole conduction to electron conduction as $V_G$ increases from zero to a positive value. Carriers are confined at the gel/BP interface with a maximum attainable sheet carrier density up to $1.2 \times 10^{14}$ cm$^{-2}$ for electrons and $0.57 \times 10^{14}$ cm$^{-2}$ for holes, corresponding to a Fermi level shift of ~0.49 eV above the conduction band minimum for electron accumulation, and of ~0.57 eV below the valence band maximum for hole accumulation (estimated by the density of states of BP from band calculations). A qualitative estimation of the carrier distribution profile of the downward band bending case under a positive $V_G$ indicates that electrons are strongly confined at the surface, while holes are distributed in the deeper bulk regime (Figs. 4f and 4g). Interestingly, the formation of the inversion layer (surface electron accumulation) on the p-type BP channel (holes in bulk) indicates that we can easily induce a vertical p-n junction structure within such an EDLT configuration. More importantly, the depth of the depletion layer and resulting p-n junction depth can be tuned with both the external perpendicular electric field (gate voltage) and the bulk carrier density in BP crystals, providing us with freedom for a rational control of the p-n junction for performance enhancement of BP photodetectors.

To this end, we vary the gate voltage $V_G$ and show the corresponding photocurrent images of the BP ring pattern under various $V_G$ (Figs. 5a-5c). With increasing $V_G$ from zero bias to 1.5 V, the maximum photocurrent under the light irradiation (1500 nm) at 0 ° polarization dramatically increases and the photocurrent generation areas enlarge, while the photocurrent under the polarized light at 90 ° does not significantly change with varying $V_G$. As indicated in Figs. 5d and 5e, the photoresponsivity at 1700 nm (0 ° polarization) can be one order of magnitude enhanced by applying a positive gate voltage, and correspondingly the ratio of the photoresponsivities between the two polarizations is greatly enhanced. The downward band bending and the resulting built-in perpendicular electric field in the BP channel in the vertical p-n junction can serve to spatially separate the photo-generated carriers (as schematically shown in the inset of Fig. 4e). Electrons move at the outmost surface while the holes move in the bulk. As the most direct result, the recombination probability between hot electrons/holes during their motion to opposite electrodes can be greatly reduced.[41, 42, 43, 44]

To generate the photocurrent, the mobile carriers will move to the electrode from the light spot center with exponential decay due to losses such as carrier recombination:

$$n(r) = n_0 e^{-\frac{r}{L_0}},$$

where $r$ is the distance from the light spot center, $L_0$ is the diffusion length and $n_0$ is the number of generated electron-hole pairs by the light shining on the sample with intensity $I_0$. Therefore two factors can directly influence the photocurrent generation: $L_0$ and $n_0$. Experimentally, if comparing the absolute distance from the BP/metal edge (20 μm in Fig. 5f) to the position at a specific photocurrent level (10 nA), one can see that a higher $V_G$ exhibits a long distance for the lateral diffusion of hot carriers, from 1.2 μm for zero $V_G$, 2.2 μm for 1.0 V, to 3.7 μm for 1.5 V. However, there is almost no change in the normalized photocurrent by the maximum value at each $V_G$. This implies that the carrier diffusion length is not significantly changed with $V_G$ and not the dominant mechanism for the photocurrent enhancement. Rather, it appears that the enhancement of $n_0$ starts to be the key point for the absolute value increase in photocurrent. Namely, as the $V_G$ increases, the tunable perpendicular electric field can separate the electrons and holes: electrons move on the surface and holes move in the bulk, which can reduce their recombination and increase the $n_0$ value.

In contrast to conventional phototransistors, photogenerated electrons and holes in our vertical p-n junction structure are separated by selectively driving them into surface or bulk layers under the built-in electric field within the EDLT, which can minimize the hot carrier recombination. Therefore, this vertical p-n junction configuration can greatly enhance the efficiency of the linear dichroism photodetector. While far from optimized, our observations provide the first demonstration that a vertical p-n junction device structure can be significant for performance enhancement in the linear dichroism photodetection scheme using BP. Key to this approach is the direct combination of the intrinsic broadband linear dichroism of BP with its high mobility semiconductor characteristics, which can enable novel integrated optical and optoelectronic device applications beyond conventional materials and approaches.

# Methods

**Device fabrication of BP photodetector**

Hall bar patterns and ring-shaped patterns with Ti/Au electrodes were fabricated on tape-cleaved BP single crystal flakes (on $SiO_2$/Si wafer), which typically have dimensions of tens of micrometers laterally and a thickness of 30-50 nm. The bulk BP crystal shows the typical p-type conduction with carrier density in the order of $10^{18}$ cm$^{-3}$. The ring pattern is designed to exclude the possibility that the observed polarization-dependent absorption might originate from anisotropic scattering from the BP/metal edge. Serving as the side gate electrode for gel gated transistors, a large-area Ti/Au pad was deposited near the BP, but electrically insulated from the BP flake. A typical EDL transistor was fabricated by spin-coating DEME-TFSI-based ionic gel[35] [DEME-TFSI, N,N-diethyl-N-(2-methoxyethyl)-N-methylammonium bis-trifluoromethylsulfonyl)-imide from Kanto Chemical Co]. Note that the ionic gel is transparent within the spectral regime from 300 nm to 3100 nm, covering the optical measurement regime (400-1700 nm). Transfer characteristics, longitudinal sheet resistance $R_s$, and Hall coefficient of Hall bar patterned devices were simultaneously measured in a liquid helium cryostat with a 14 T superconducting magnet.

**Optical measurements and scanning photocurrent microscopy**

Optical reflectance and transmittance spectral measurements were carried out under ambient conditions at room temperature for samples on transparent quartz substrates. The spectra in the visible regime (1.5-3.0 eV) were obtained with a confocal scanning microscope (Nikon LV-UDM) equipped with a halogen lamp, a grating spectrometer and a CCD camera. The detection area was with a spot diameter of 10 μm. For reflection measurements in the infrared, we used a Nicolet 6700 Fourier-transform infrared (FTIR) spectrometer equipped with a Continuum XL microscope (Thermo Electron Corp.). The IR light source was focused by a 15× Cassegrain objective with angles ranging from 16 to 35.5 degree with respect to the sample normal. A variable knife-edge aperture located within the image plane is used to define the sample collection area, which is fixed in size to ~200 μm × 200 μm. The spectral range was 650-6500 cm$^{-1}$ (0.08~0.8 eV) with a resolution of 4 cm$^{-1}$. All spectra are the average of 32 scans. The reflectance spectra were plotted as ($R_{sample}/R_{Au}$) ×100%, where $R_{sample}$ was collected from an area with a sample present and $R_{Au}$ was collected from an adjacent Au pad as a total reflection mirror.

In photocurrent measurements, optical radiation of a well-defined wavelength is selected from a supercontinuum Fianium white light laser source using acousto-optic tunable filters. The laser beam passes through a linear polarizer and a half-wave plate and is then focused on the detector using a microscope objective (50×, 0.42 NA). The laser beam is modulated with a mechanical chopper (1 KHz) and the modulated photocurrent signal is amplified and detected using a lock-in technique. The photodetector device is mounted on a piezoelectric stage and a photocurrent image of the detector is obtained by raster scanning the laser beam over it. To investigate the time response of the detector, we illuminate it instead with an electrically modulated diode laser at 780 nm (Toptica DL-100), driven by a square wave function generator. After passing through a preamplifier, the photodetector signal is recorded in time by a digital oscilloscope.

**ARPES and *ab initio* band calculations**

High-resolution ARPES experiments for cleaved BP samples were performed at the Synchrotron Radiation Light Source Beamline 5-4 under a base pressure better than $3 \times 10^{-11}$ Torr, using incident photon energies of 19.5 eV with energy resolution of 12.5 meV and angular resolution of 0.2°. Electronic structure calculations were carried out by using the Vienna Ab-initio Simulation Package (VASP)[45, 46] based on experimental lattice constants and optimized internal parameters with accuracy smaller than 0.01 eV/A. Perdew-Burke-Ernzerhof-type generalized gradient approximation[47,48] with MBJ correction was used to obtain an accurate semiconducting gap. The k-mesh for self-consistent calculations is $12 \times 12 \times 16$, and the cut-off energy is fixed to 400 eV.

**Author Contribution**

H.T.Y., H.Y.H. and Y.C. conceived and designed the experiments. H.T.Y. performed sample fabrication and transport measurements. H.T.Y., X.G.L., F.A., A.G.C. and M.B. performed optical measurements. W.L. and Z.X.S. performed ARPES measurement. G.X., B.L. and S.C.Z. performed DFT calculations and theoretical analyses. X.G.L performed the band bending calculation. J.S. performed TEM analysis. G.J.Y and X.H.C. grew the high quality BP crystals. Y.H., M.B., Z.X.S., S.C.Z., X.H.C., H.Y.H. and Y.C. supervised the project and all authors contributed to data discussions. H.T.Y. wrote the manuscript with input from all authors.

**Acknowledgements**

This work was supported by the Department of Energy, Office of Basic Energy Sciences, Division of Materials Sciences and Engineering, under contract DE-AC02-76SF00515. B.L., G.X., H.Y.H. and S.C.Z. also acknowledge FAME, one of six centers of STARnet, a Semiconductor Research Corporation program sponsored by MARCO and DARPA.

**Author Information**

The authors declare no competing financial interests. Correspondence and requests for materials should be addressed to hyhwang@stanford.edu (H.Y.H.), yicui@stanford.edu (Y.C.).

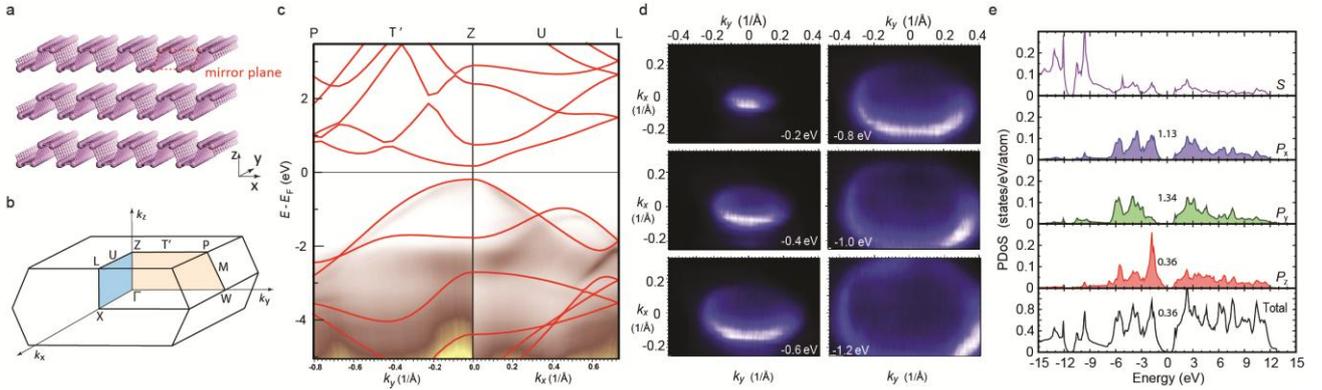

**Figure 1 Anisotropic electronic structure of layered black phosphorus. a,** and **b,** Layered crystal structure of black phosphorus and the schematic of its 3D Brillouin zone. The BP lattice shows a novel in-plane atomic buckling along the *x*-axis, resulting in a two-fold anisotropy along *x* and *y* directions, with a direct influence on its band structure and the optical selection rules. The parallelogram plane (red color) in **a** reflects the mirror reflection symmetry in BP structure. **c,** Band structure of cleaved bulk BP, obtained by ARPES measurements (color mapping) and from band calculations (red lines), showing the anisotropic band dispersion and effective mass between Z-P and Z-L directions. **d,** Constant-energy contours show the evolution of the anisotropic band dispersion at different energies from -0.2 eV to -1.2 eV. The energies here are all with respect to the Fermi level. **e,** Partial density of states (PDoS) of BP for *s*, $p_x$, $p_y$, $p_z$ orbitals. The number in each panel indicates the gap size for each PDoS, where the $p_z$ orbital along the *z* crystal axis contributes all of the band dispersion near the band edge and practically dominates the size of band gap of bulk BP.

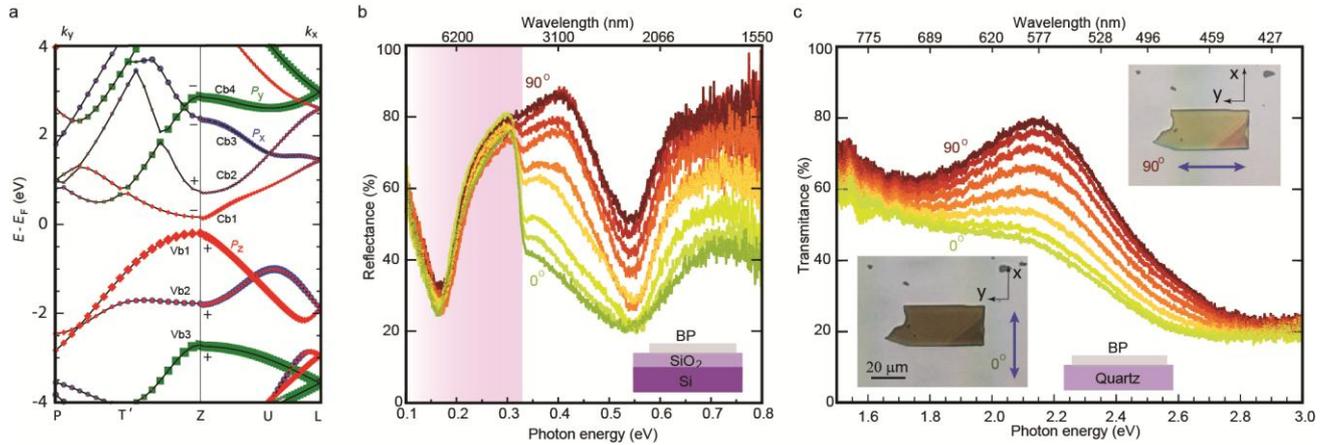

**Figure 2 Optical selection rules and broadband linear dichroism in black phosphorus. a,** Orbital component analysis of bulk BP band dispersion. In the Z-L direction, only *x*-direction polarized light can be absorbed. In the Z-P direction, light with both *x* and *y* polarizations can be absorbed. "+" and "−" signals represent the parity of the bands. Red, blue and green colors represent the band dispersion from the $p_z$, $p_x$, $p_y$ orbitals. **b,** Light polarization dependence of the reflection in the infrared spectral regime, showing around 50% variation in reflection along two perpendicular directions for energies above the band gap (purple shadow area). **c,** Polarization dependence of the transmission of visible light. In both spectra, the incident light is linearly polarized with directions from the *x* to the *y* crystal axis (from green/0° to dark-red/90°) in 15° steps. The insets in **c** show optical images with incident light along two perpendicular directions (blue arrow represents light polarization direction).

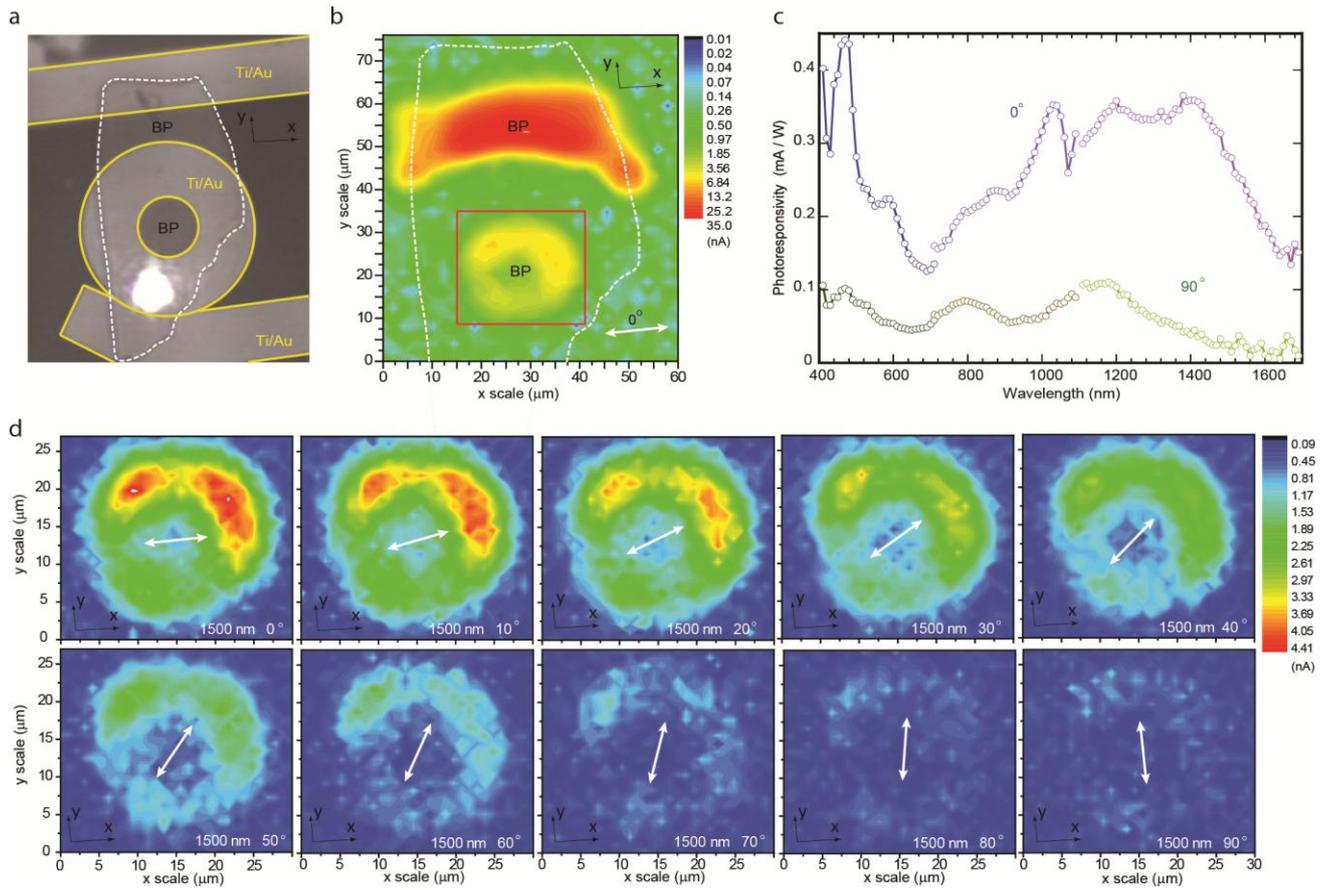

**Figure 3 Black phosphorus photodetector with broadband and polarization sensitivity. a,** Optical image of BP photodetector with ring-shaped photocurrent collector. The areas indicating by yellow lines are Ti/Au electrodes and the area circled by white line is the BP flake. In contrast to a "knife"-edge metal electrode, the isotropic round shape photocurrent collector can remove the linear dichroism which might originate from a straight metal edge. **b,** Corresponding photocurrent microscopy images of the device shown in **a**, with illumination at 1500 nm and polarization along the *x* direction (white arrow). Further detailed investigations on the photocurrent generation focus on the BP inside the inner ring. **c,** Polarization dependence of the photoresponsivity under illumination from 400 to 1700 nm, where the polarization angle of 0° corresponds to the *x* crystal axis while 90° corresponds to *y* crystal axis. **d,** Photocurrent microscopy images of the BP inside the inner ring under illumination at 1500 nm with different light polarizations (white arrows reflect the polarization directions. As shown in Supplementary Figs. S5 and S6, similar results are observed under broadband illumination, at least within the spectral regime in our experiment (400 – 1700 nm).

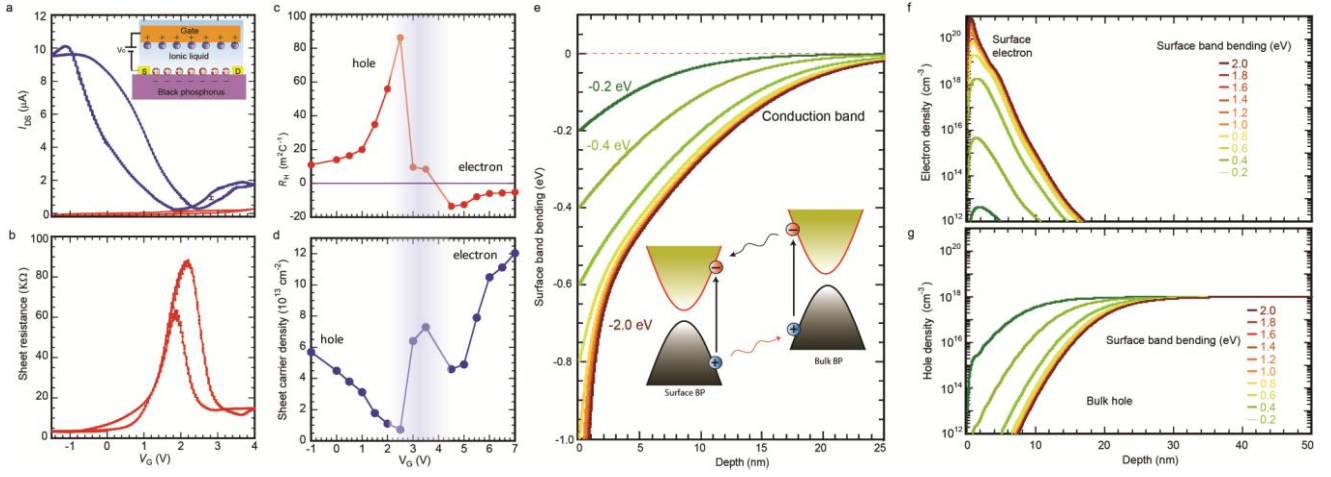

**Figure 4 Ambipolar operation and the vertical p-n junction in BP transistors. a,** Ambipolar transfer characteristics in ionic gel gated electric-double-layer transistors based on a cleaved BP flake with a thickness of ~30 nm. Inset: Schematic structure of a typical ionic gel gated BP transistor. By applying a gate voltage $V_G$ to the lateral Au gate electrode, ions in the gel are driven to the BP surface, forming a perpendicular electric field at the BP surface. The blue curve is the source-drain current and the red curve is the corresponding gate current, both measured at 230 K. **b,** Sheet resistance as a function of gate voltage $V_G$, indicating an ambipolar behavior similar to graphene transistors. **c,** and **d,** Hall coefficient and sheet carrier density obtained from Hall effect measurements. The maximum attainable sheet carrier densities of $0.57 \times 10^{14}$ cm$^{-2}$ for holes and $1.2 \times 10^{14}$ cm$^{-2}$ for electrons indicate the existence of surface band bending on the BP side. **e,** Self-consistent Poisson-Schrödinger calculations for surface band bending in ionic gel gated BP transistors. Green color represents a weaker downward band bending (-0.2 eV) while dark-red color represents a stronger band bending (-2.0 eV). Inset: schematic diagram of the relative movement of the photogenerated electron/hole pairs in the built-in electric field in a vertical p-n junction. **f** and **g,** Carrier distribution profile in ionic gel gated BP transistors obtained from Poisson-Schrödinger calculations for different band bending (as indicated by the labels), demonstrating the accumulation of electrons on BP surface. Here we use the bulk hole density of $1 \times 10^{18}$ cm$^{-3}$. A higher bulk carrier density would give a narrower p-n junction (space charge region in the vertical direction) owing to a smaller Thomas-Fermi screening length.

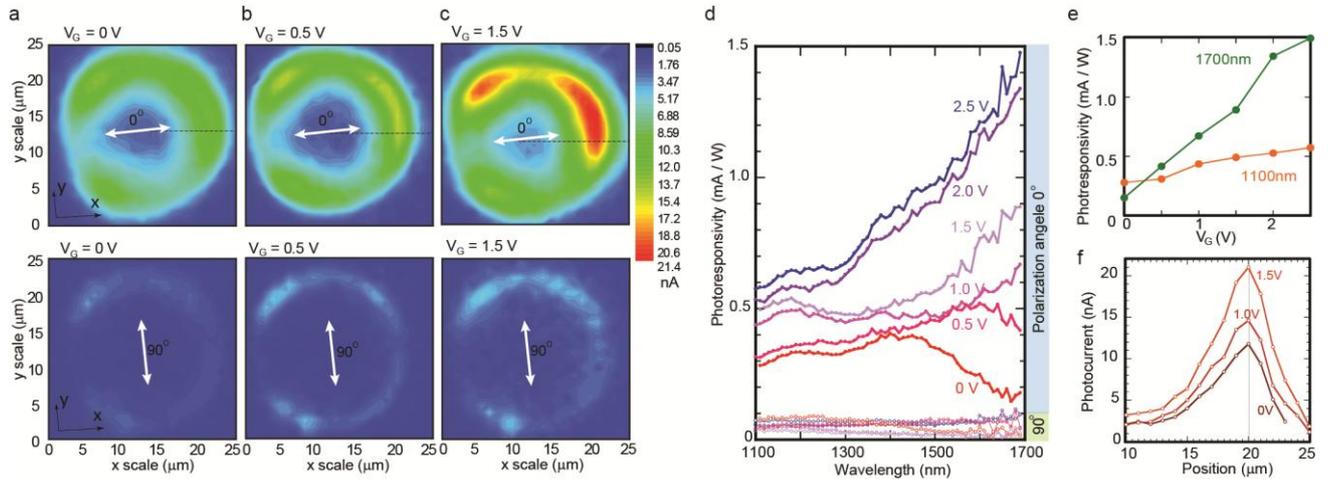

**Figure 5 Gate enhancement of the linear dichroism detection by a vertical p-n junction in a BP electric-double-layer transistor. a**-**c**, Photocurrent microscopy images of the BP device with illumination at 1500 nm for two perpendicular polarizations under gate biases $V_G = 0$ V (**a**), 0.5 V (**b**) and 1.5 V (**c**). A polarization angle of 0 ° corresponds to the *x* crystal axis while 90 ° corresponds to the *y* crystal axis, as indicated by white arrows. **d**, Gate enhanced linear dichroism detection in BP photodetector as demonstrated by the dependence of the photoresponsivity for incident light polarized along the *x* and *y* crystal axes. **e**, Gate dependent photoresponsivity at two different wavelengths (1100 nm and 1700 nm) for light polarization along the *x* axis. **f**, Gate dependent photocurrent profile from the center of the ring to the metal edge along the black dash line shown in panels **a**-**c**, for light polarization along the *x* axis (1500 nm).

# Broadband Linear-Dichroic Photodetector in a Black Phosphorus Vertical p-n Junction

Hongtao Yuan,[1,2] Xiaoge Liu,[1] Farzaneh Afshinmanesh,[1] Wei Li,[1] Gang Xu,[1] Jie Sun,[1] Biao Lian,[1] Alberto G. Curto,[1] Guojun Ye,[3] Yasuyuki Hikita,[1,2] Zhixun Shen,[1,2] Shou-Cheng Zhang,[1,2] Xianhui Chen,[3] Mark Brongersma,[1,2] Harold Y. Hwang,[1,2*] Yi Cui[1,2*]

[1]*Geballe Laboratory for Advanced Materials, Stanford University, Stanford, California 94305, USA*
[2]*Stanford Institute for Materials and Energy Sciences, SLAC National Accelerator Laboratory, Menlo Park, California 94025, USA*
[3]*Hefei National Laboratory for Physical Science at Microscale and Department of Physics, University of Science and Technology of China, Hefei, Anhui 230026, China*

**Supplementary Information**

**Supplementary Section 1: Determination of crystal orientation for linear dichroism and anisotropic light absorption in black phosphorus**

Although the layered structure of black phosphorus is similar to other 2D materials by stacking the monolayer unit along the *z*-axis (Fig. 1a), the puckered in-plane structure in a BP monolayer, having mirror reflection symmetry and inversion symmetry (Supplementary Fig. S1a), serves as the origin of the anisotropic properties between *x* and *y* directions. Therefore it is of great importance to confirm the crystal orientation of the BP samples. Interestingly, when we cleave the BP sample with the mechanical exfoliation method, it was found that the obtained flakes (cleaved from needle-like bulk crystals) always have a long stripe shape with two sharp parallel edges (Supplementary Fig. S1b), indicating a preferential cleaving orientation for BP. In order to determine the relative orientation of cleaved samples, we performed high resolution transmission electron microscopy (HRTEM) imaging on more than ten samples. The typical HRTEM image and the Fast Fourier transform (FFT) shown in Supplementary Fig. S1c and its inset are used for crystal orientation and lattice parameter analysis. The interplanar distance parallel to the long side is 2.2 Å corresponding to the [020] orientation (Supplementary Fig. S1c), which was calculated based on the distance between two centrosymmetric dots vertical to the long side. Thus, we can clearly establish that the crystal direction with long edges is the *y* axis of the cleaved BP while the crystal direction with the short edges is the *x* axis as indicated in Supplementary Fig. S1d.

---

[*] Corresponding author: <u>hyhwang@stanford.edu</u>, <u>yicui@stanford.edu</u>.

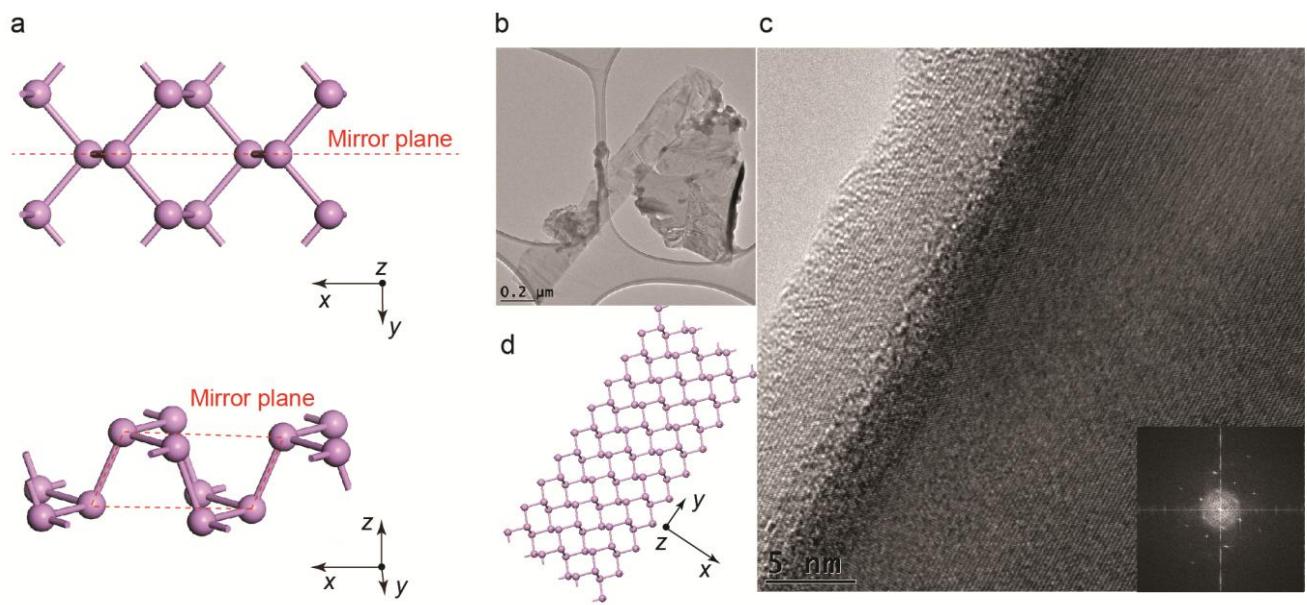

**Supplementary Figure S1. a,** The mirror plane and the mirror reflection symmetry in the puckered BP structure. **b,** Cleaved BP flakes on a TEM grid, with a long stripe shape and two long parallel edges. **c,** High-resolution transmission electron microscopy image for the preferential crystal orientation in cleaved BP. Inset: Fast Fourier transform pattern of **c**. The long stripe edge direction is confirmed to be BP's *y* direction, as schematically shown in **d**.

**Supplementary Section 2: Anisotropic effective mass in black phosphorus and anisotropic electronic transport in BP EDLTs**

Theoretical band calculations can establish the band dispersion anisotropy of bulk BP, especially the difference of the effective mass along two anisotropic directions. Supplementary Table I shows the anisotropic effective mass of free carriers along three perpendicular directions, obtained from our *ab initio* band calculations.[1] One can see that near Z point in the momentum space, the valence band (also conduction band) along the Z-L direction disperses rather strongly compared to that of the Z-P direction, indicating a smaller effective mass along the Z-L direction. A fit of these bands using the nearly-free electron model gives the small effective carrier masses along the Z-L direction with values of 0.11 $m_0$ for electrons and 0.12 $m_0$ for holes, whereas these along Z-P are much larger with the value of 1.30 $m_0$ for electrons and 0.81 $m_0$ for holes. More interestingly, in the third direction (the *z* direction), the carriers' effective masses along Z-Γ direction are found also relatively small, providing advantages for potential device applications with vertical structures such as solar cells.

Supplementary Table I. Calculated effective mass along the anisotropic directions in bulk BP

| Effective mass | electron ($m_0$) | hole ($m_0$) |
| --- | --- | --- |
| $m_x$ | 0.11 | 0.12 |
| $m_y$ | 1.3 | 0.81 |
| $m_z$ | 0.18 | 0.37 |

As discussed in the main text, light absorption spectra with linear dichroism in BP make an optical determination of the crystalline orientation possible, and thus provide a way to fabricate electronic devices along a certain direction.[1, 2] Supplementary Figures S2a-e show the ambipolar transfer characteristics of a Hall-bar patterned BP EDLTs with gel-gating (inset of Supplementary Fig. S2a), measured along different crystal directions where the *x* (Supplementary Fig. S2b) and *y* (Supplementary Fig. S2e) axis with most and least absorption can be easily distinguished by using optical microscopy with a linear polarizer. Within a fixed $V_G$ scale, the maximum attainable source-drain current ($I_{DS}$) along the *x* (pins 1-4) and *y* axis (pins 4-2) gives quite different values, roughly around a factor of 5 for the electron side and 2 for the hole side, which reflects the anisotropy in momentum space for both electrons and holes[3]. Note that the relatively high mobility of our BP devices in the range of a few hundreds to a thousand cm$^2$/V s also implies that the photo-excited carrier can have a long mean free path.

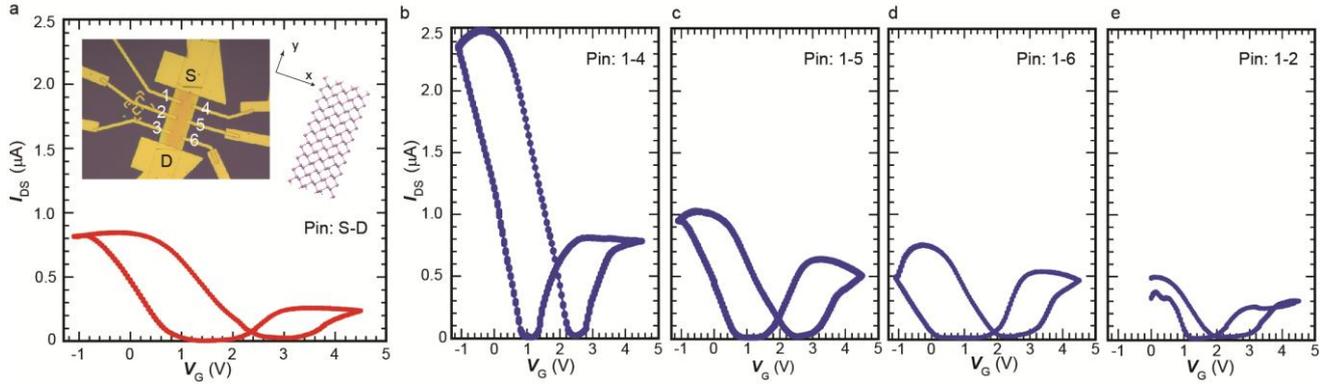

**Supplementary Figure S2. a,** Hall-bar patterned BP electric-double-layer transistor (EDLT) with ionic gel gating and the transfer characteristics measured between source and drain electrodes. **b-e,** Transfer characteristics of the BP EDLT, measured between pins 1-4, 1-5, 1-6 and 1-2, reflecting the intrinsic anisotropic transport in BP.

**Supplementary Section 3: Linear dichroism for light absorption in black phosphorus**

To confirm the linear dichroism in BP, visible light reflection, transmission, and absorption spectra were performed by using an inverted confocal microscope (Nikon LV-UDM microscope). Infrared reflection spectra where obtained through Fourier transform infrared spectroscopy (FTIR). Schematic optical setups for FTIR and inverted microscope are shown in Supplementary Figs. S3a and S3b. The reflectance spectrum was plotted as $(R_{sample}/R_{metal}) \times 100\%$, where $R_{sample}$ was collected from an area with a sample present on $SiO_2$/Si substrate and $R_{metal}$ was collected from an adjacent Au or Al pad as a total reflection mirror. The transmittance spectrum were plotted as $(T_{sample}/T_{quartz}) \times 100\%$, where $T_{sample}$ was collected from an area with a sample present on quartz glass and $T_{quartz}$ was collected from an adjacent area without BP flakes. Absorption spectra were estimated as (1-Reflection-Transmission)$\times 100\%$, since our sample was much larger than the light spot and had an atomically smooth surface, such that scattering was negligible.

Polarization dependent reflection, transmission and absorption measurements on cleaved BP flakes were performed to confirm linear dichroism response in BP flakes with visible light. Similar to the related measurements shown in the main text, as we change the light polarization from the *x* direction (green spectra in Supplementary Figs. S3c-e) to the *y* direction (red spectra in Supplementary Figs. S3c-e), the reflectance and transmittance start to increase while the absorbance starts to decrease. With the light polarized along the *y* direction, the absorption can reach zero within some wavelength regime and the flakes can be much more transparent with very low absorption. A video in the Supplementary Information clearly indicates the gradual changes of the transmittance during continuous rotation of the linear polarizer.

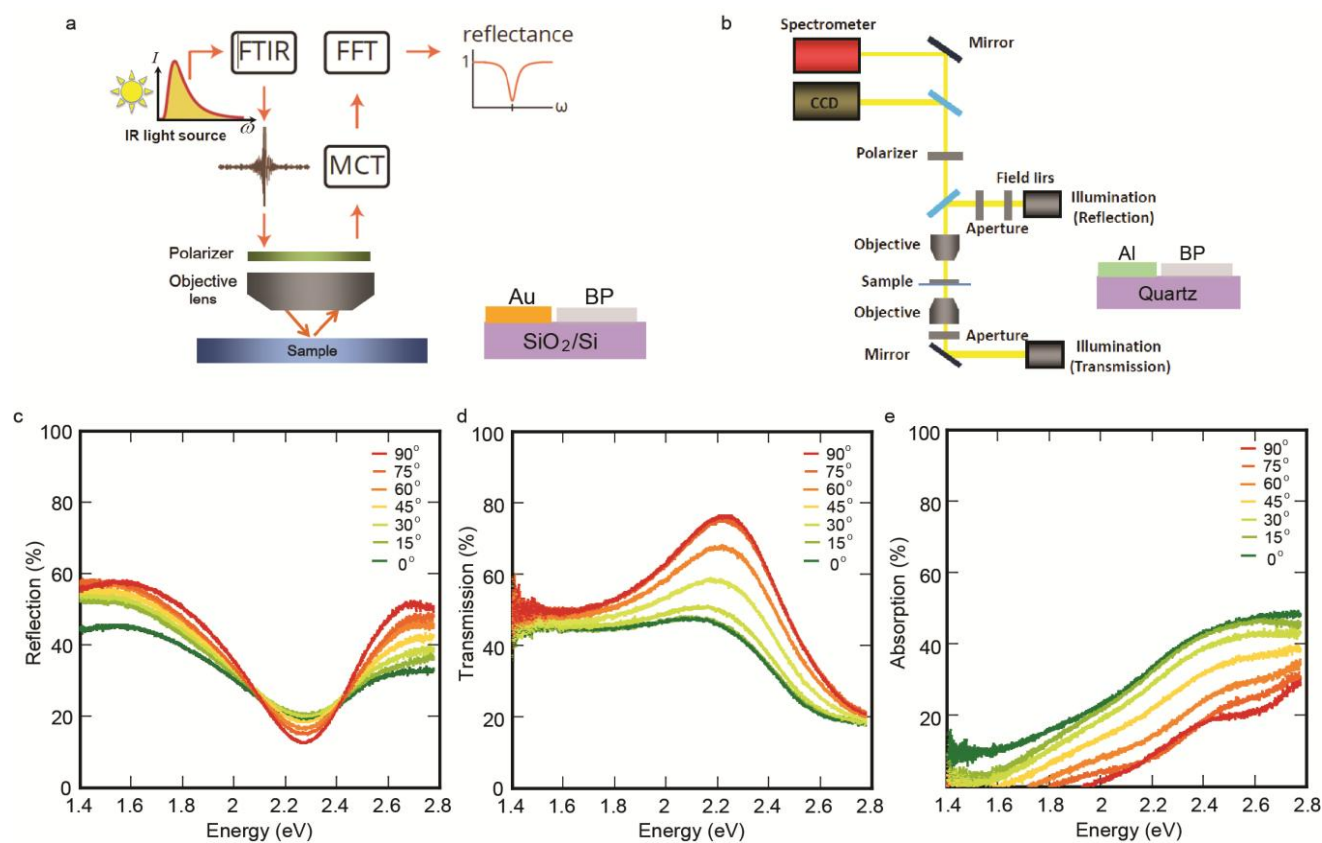

**Supplementary Figure S3. a,** and **b,** Schematic optical measurement setup with infrared and visible light. Light polarization dependent reflection (**c**), transmission (**d**) and absorption (**e**) for cleaved BP flakes. Green color represents light polarization along the *x* direction (0 °polarization) and red for the *y* direction (90 °polarization)..

Here we give the theoretical prediction of a broadband polarization-sensitive absorption spectrum up to 3.0 eV. The full optical transition behaviour of the material can be obtained based on the optical transition rules and the band structure calculation shown in Fig. 2a of the main text. As the photon energy increases, more and more bands will contribute to the optical transition. We are able to determine the mirror reflection symmetry ($y \rightarrow -y$) in the $k_y = 0$ plane of all the low energy bands, as shown in Table II. According to the optical selection rules, transitions between two bands of the same (opposite) mirror (*y*) symmetries will favour the absorption of *x*-polarized (*y*-polarized) light. From Fig. 2a in the main text, it is clear that the two bands Vb3 and Cb4 (both have negative mirror symmetry) are not involved until the photon energy is above around 3.0 eV (the transitions between Vb1 ↔ Cb4 and Vb3 ↔ Cb1 begin to take place). Therefore, in a broadband of photon energies up to around 3.0 eV, the optical transition only happens between bands with the same mirror (*y*) symmetry, and the absorption of *x*-polarized light will be favoured. As a direct result, the anisotropic absorption will be greatly suppressed with the photon energy large than 3.0 eV.

Supplementary Table II. Mirror (*y*) symmetry of bands in the $k_y = 0$ plane determined by calculations.

| Band | Vb3 | Vb2 | Vb1 | Cb1 | Cb2 | Cb3 | Cb4 |
|---|---|---|---|---|---|---|---|
| Mirror (*y*) | - | + | + | + | + | + | - |

Particularly, we have directly calculated the optical transition amplitudes between parity allowed bands at the high symmetry Z point, and the results are shown in Table III. It is found that there is only absorption of *x*-polarized light between Vb1 ↔ Cb1 at Z point. This shows that the light absorption is largely contributed by the transition between Vb1 ↔ Cb1. The contributions from the other bands only arise in regions of the Brillion Zone with a lower symmetry.

Supplementary Table III. Optical transition amplitudes (arbitrary unit) of x and y polarized light at Z point.

| Pair of bands | Vb1 ↔ Cb1 | Vb1 ↔ Cb3 | Vb1 ↔ Cb4 | Vb2 ↔ Cb1 | Vb3 ↔ Cb1 |
|---|---|---|---|---|---|
| *x*-amplitude | 0.247 | 0 | 0 | 0 | 0 |
| *y*-amplitude | 0 | 0 | 0 | 0 | 0 |

**Supplementary Section 4: Polarization-sensitive photocurrent for linear dichroism detection in Hall-bar patterned black phosphorus device.**

As discussed in the main text, the scanning photocurrent microscopy measured over the Hall-bar patterned BP device clearly shows the evolution of photocurrent as a function of the polarization angle of the linear polarizer. With the laser spot scanning near the metal electrode edges (the yellow square in Supplementary Fig. S4c), the photocurrent with incident light at 1500 nm wavelength polarized along the *x* crystal axis (defined as 0° polarization) shows a strip shape and it is much stronger than that with incident light polarized along the *y* crystal axis (defined as 90° polarization), establishing the polarization dependent absorption and resulting detection for the linear dichroism of incident light. Note that the photocurrent flowing directions at two electrodes are opposite to each other, as discussed later.

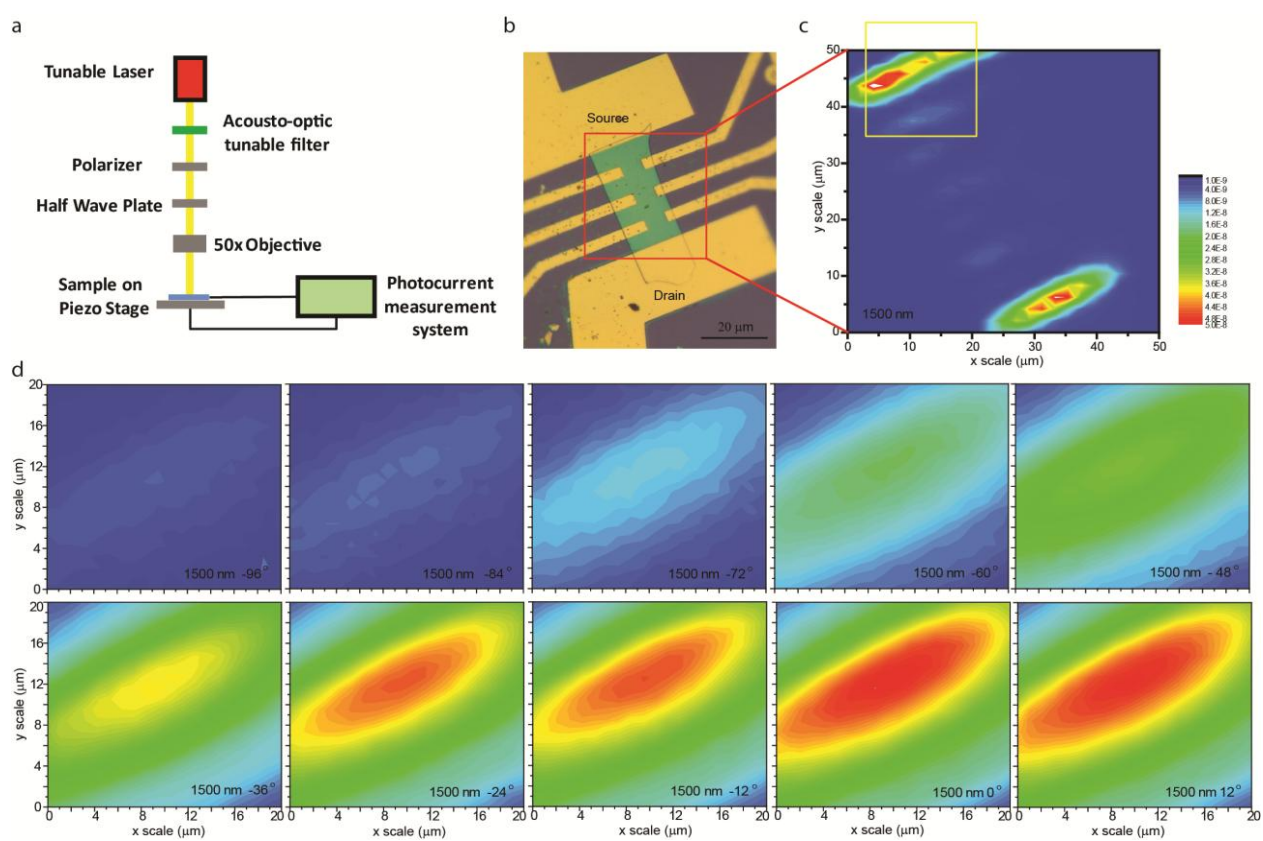

**Supplementary Figure S4. a,** Schematic of the scanning photocurrent microscope. **b,** Hall bar patterned BP device. **c,** Large area photocurrent mapping for the red square area shown in **b**. **d,** Polarization-dependent photocurrent images for linear dichroism detection for the yellow square area shown in c.

**Supplementary Section 5: Broadband polarization-sensitive photocurrent for linear dichroism detection in infrared and visible light regime.**

Since the observed two-fold polarization dependent photocurrent can be affected by extrinsic anisotropic scattering of photo-excited carriers from the geometry of the metal-electrode edge, we designed a "ring" shape metal electrode as the photocurrent collector (Fig. 3a), aiming to collect photogenerated carriers isotropically to eliminate the influence from the electrode edge. Similarly to the result shown in the Hall-bar pattern devices in the previous section, here the photocurrent with 0 ° polarization is also much larger than that for 90 ° polarization (Fig. 3c), directly reflecting the intrinsic polarization dependent photoresponse from BP itself. Additional measurements on the light polarization dependent photocurrent mapping under various wavelengths from 420 nm to 1500 nm are given in Supplementary Figs. S5 and S6, indicating a broadband detection for linear dichroism.

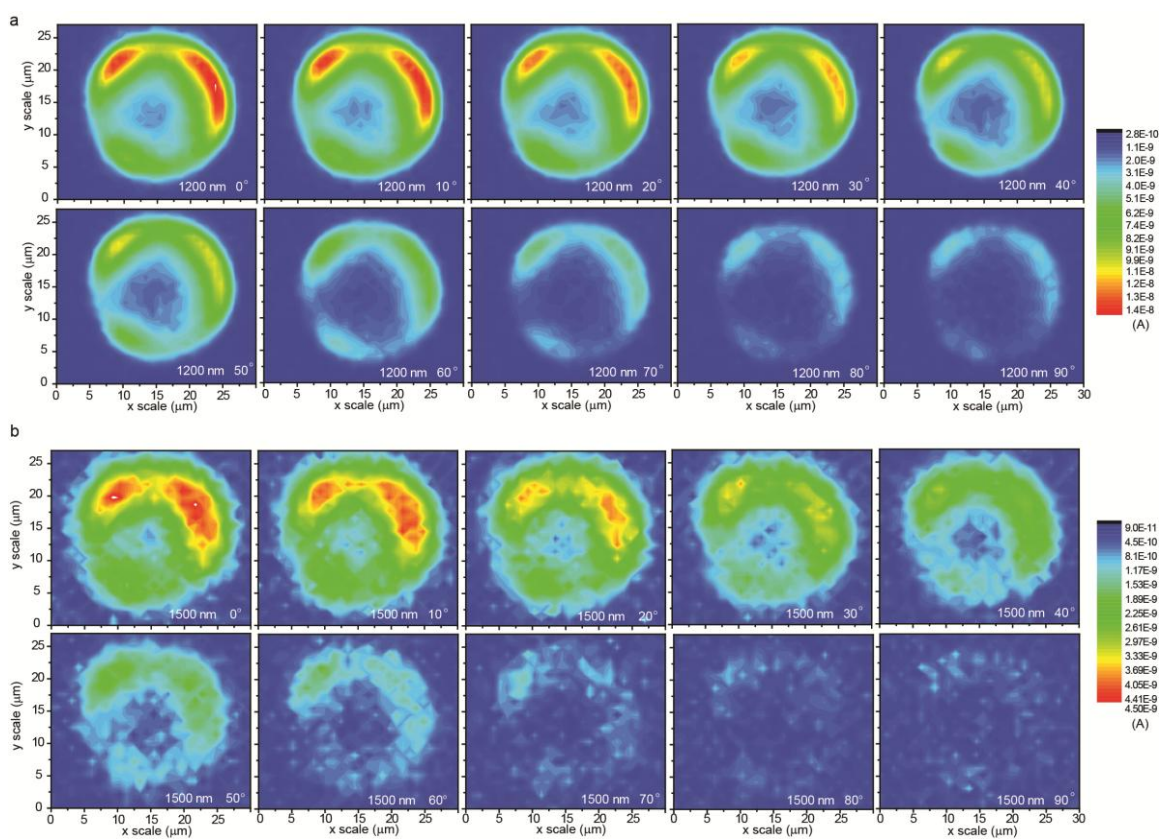

**Supplementary Figure S5.** Broadband polarization-sensitive photocurrent for linear dichroism detection in ring-shaped patterned black phosphorus device in infrared regime (1200 nm in **a,** and 1500 nm in **b**). The zero angle represents light polarization along the *x* crystal direction while 90 ° represents light polarization along the *y* crystal direction.

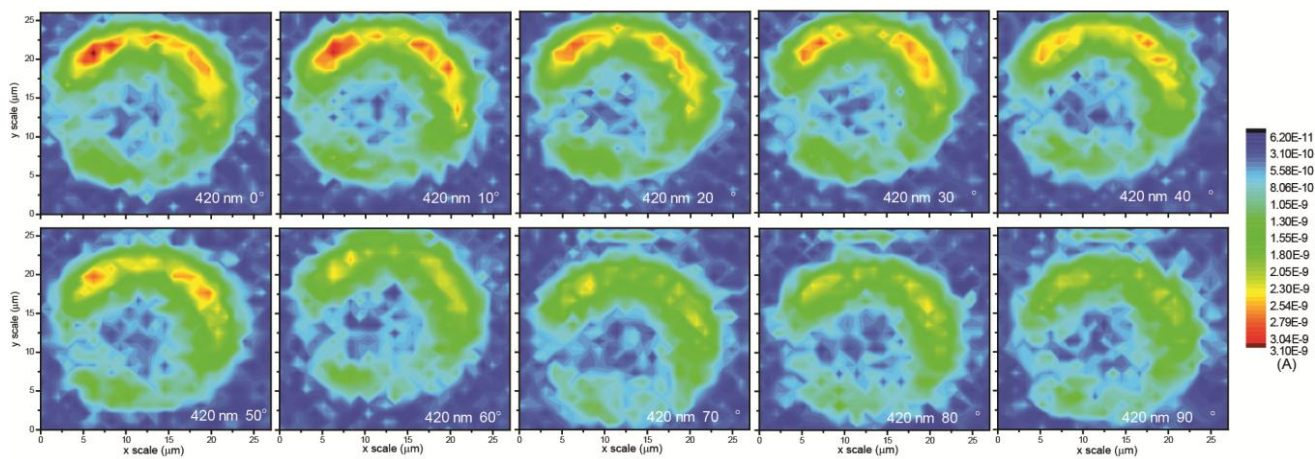
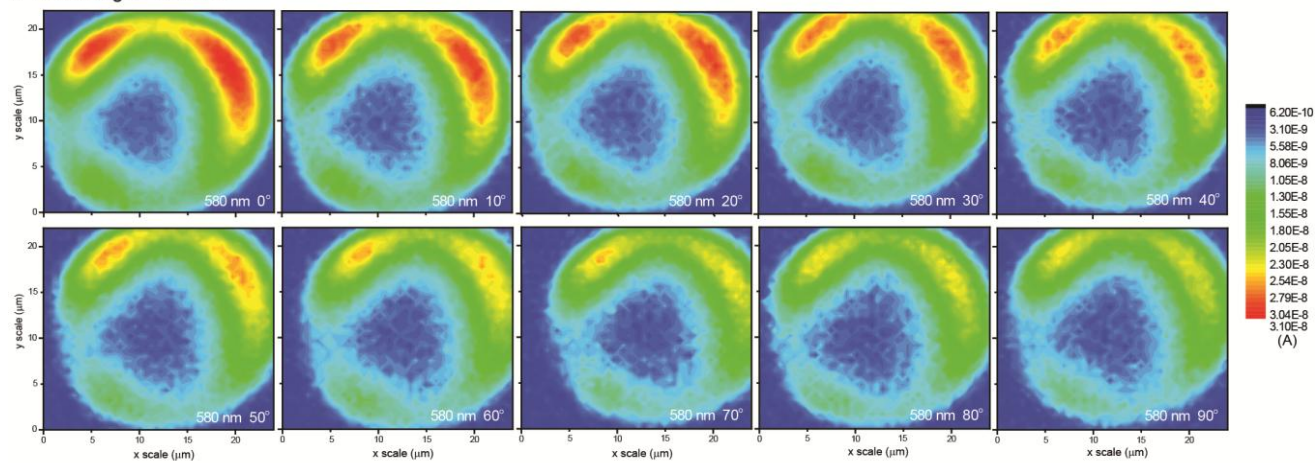
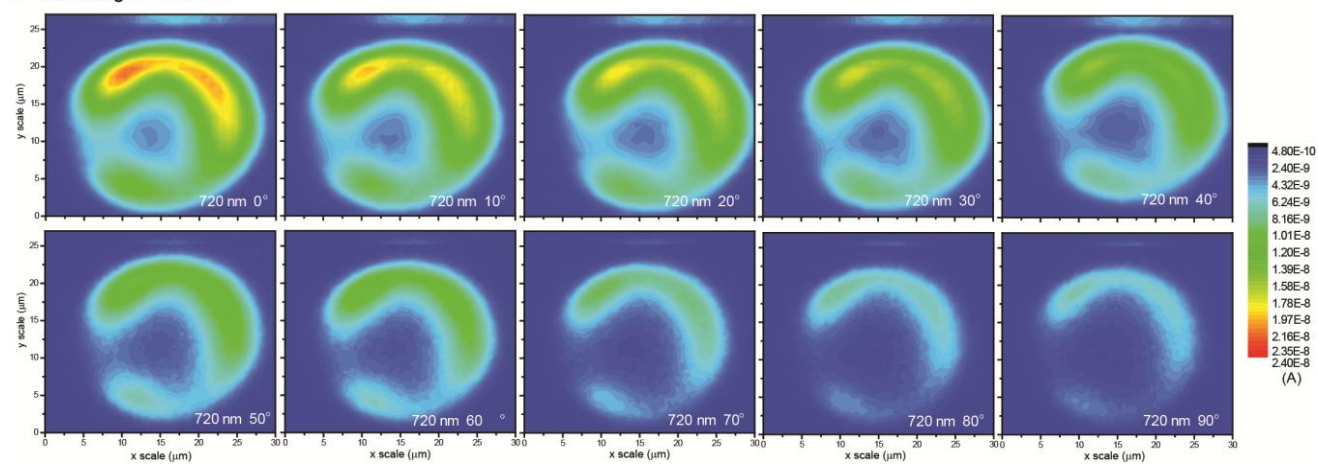

**Supplementary Figure S6.** Broadband polarization-sensitive photocurrent for linear dichroism detection in Ring-shape patterned black phosphorus device in visible light regime (**a,** 420 nm; **b,** 580 nm and **c,** 720 nm). The zero polarization 0 ° represents light polarization along the *x* crystal direction while 90 ° represents light polarization along the *y* crystal direction.

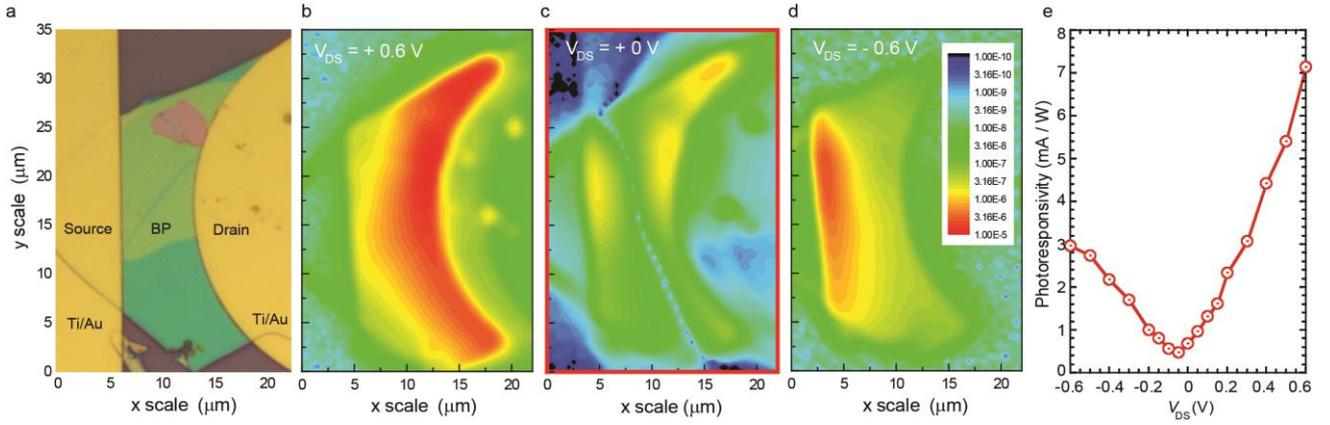

**Supplementary Figure S7.** Photoresponse in BP photodetector as a function of $V_{SD}$. **a,** Optical image of the device configuration. **b-d,** Photocurrent $|I_{ph}|$ mapping images for the $V_{SD}$ at +0.6 V, 0 V, -0.6 V. **e,** The maximum photoresponsivity as a function of $V_{SD}$. Laser excitation has a wavelength of 780 nm with power 1.35 mW.

## Supplementary Section 6: Photocurrent generation mechanism in black phosphorus photodetector: from the photo thermoelectric effect to the photovoltaic effect

As the most important photocurrent generation mechanism, photothermal currents driven by a temperature gradient ΔT always include both photo thermoelectric currents (with equal charge carrier and lattice temperatures) and hot carrier transport (where the charge carrier temperature is higher than the lattice temperature). In a photo thermoelectric process, if the generated photocurrent equals σSΔT and the power of the incident laser is proportional to $(\kappa_e + \kappa_{ph})\Delta T$, the thermoelectric photoresponsivity is thus proportional to $\sigma S/(\kappa_e + \kappa_{ph})$, where σ is the electrical conductivity, S is the Seebeck coefficient and $\kappa_e$ ($\kappa_{ph}$) are the electronic (phononic) thermal conductivities. A high electrical conductivity together with a low in-plane lattice thermal conductivity in BP indicates that this layered semiconductor is naturally an ideal system for photo thermoelectric current generation. One of the important features for the photo thermoelectric current is the unique spatially dependent photoresponse, namely the localized photoresponse near the electrodes and a zero-photocurrent around the center of the channel. In contrast, the photovoltaic current generation mechanism is based on an electric separation of photogenerated electron–hole pairs by the lateral built-in electric fields at junctions and can be controlled with an applied source–drain bias ($V_{SD}$) between two electrodes. Since the photocurrent direction depends only on the direction and the magnitude of the lateral built-in electric field, not on the overall doping level, it switches sign when switching the polarity of the applied bias $V_{SD}$.

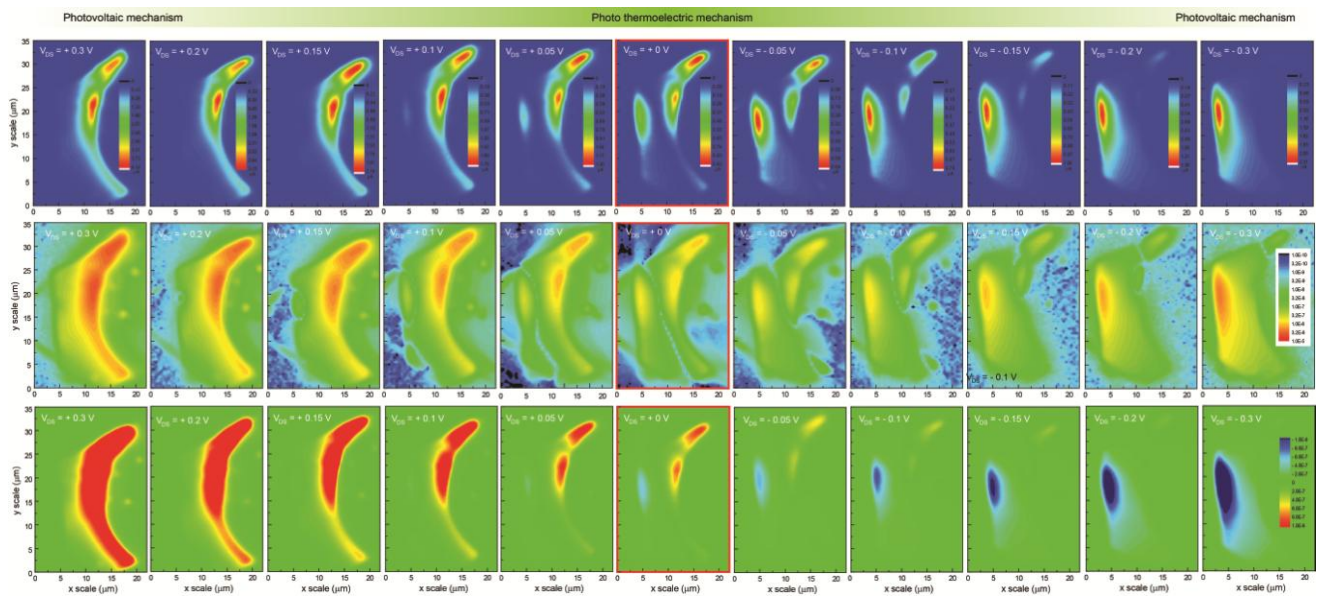

**Supplementary Figure S8.** Photocurrent generation mechanisms in BP photodetector are determined from $V_{DS}$ bias-dependent photocurrent maps. From left to right, $V_{DS}$ decreasing from +0.3 V to -0.3 V. The first row is the $V_{DS}$ dependent photocurrent $|I_{ph}|$ in linear scale, the second row is the $V_{DS}$ dependent photocurrent $|I_{ph}|$ in logarithmic scale and the third row is the photocurrent direction reflected by $|I_{ph}|\sin(\phi)$, where $\phi$ is the phase of the current measured by the AC lock-in amplifier. The definition of the positive current is the flow from the source electrode to drain electrode. Red color indicates positive photocurrents from source to drain while blue color indicates negative photocurrents from drain to source. Laser excitation has a wavelength of 780 nm and power 1.35 mW.

Supplementary Figures S7 and S8 show the photoresponse in BP photodetector as a function of $V_{SD}$. Note that for $V_{SD}$ = 0 V (Supplementary Fig. S7c) there is a clear zero crossover of the photocurrent around the central area between the source and drain electrodes. Such a zero crossover exists with a small applied $V_{SD}$, independent of the voltage polarity, which directly proves the photo thermoelectric mechanism for the photocurrent generation in BP photodetector at low $V_{SD}$. As indicated in more detail in Supplementary Figure S8, with increasing $V_{SD}$, depending on the voltage polarity, this narrow zone with zero photocurrent will gradually move to one of the electrode and finally disappear there with $|V_{SD}| > 0.2$ V. At a direct result, the photocurrent in the whole channel has the same flowing direction and this intensity/gradient of the photocurrent increases with the $V_{SD}$ value (Supplementary Fig. S7e). This can be ascribed to the photovoltaic effect originating from the built-in electric field near the metal electrodes rather than the photo thermoelectric effect under pulsed excitation since there is no zero-photocurrent crossing along the current channel. In short, the photo thermoelectric effect dominates the linear dichroism photodetection at zero or low DC voltages ($|V_{SD}| <$ ~0.15 V) while the photovoltaic effect starts to dominate the photocurrent generation at higher $|V_{SD}|$.

**Supplementary Section 7: Ultrafast photoresponse in BP linear dichroic photodetectors**

As mentioned in the main text, BP is not only a good light absorber but also a good carrier transport material owing to its small effective mass and thus high mobility. Actually this advantage can be directly seen in the photoresponse speed of the BP photodetector devices. As shown in Supplementary Fig. S9, the photodetector can easily respond to the pulsed laser for both switch-on and switch-off processes with a speed faster than a value of 40 micro-seconds (40 μs), which actually is limited by the detectable time-response of the preamplifier in our measurement setup (as confirmed by using a commercial GHz ultrafast photodetector instead of the BP detector). This implies that the real photoresponse time will be even faster than 40 μs. Compared to all photodetectors based on non-graphene 2D materials reported so-far, with the time response normally in millisecond (ms) order, such a speed of < 40 μs in the BP photodetector have been good enough for applications.

Here we numerically estimate the real photoresponse time of the ultrafast BP photodetectors using the following simple model. Since the response time is in principle limited by the carrier travel time through the black phosphorus channel region between source and drain electrodes (channel length $L$), the slowest case would be the light shining on one end of the channel and the photogenerated carriers need to travel though the entire channel region. A simple estimation would be:

$$t = \frac{L}{v} = \frac{L}{\mu E} = \frac{L^2}{\mu V_{SD}} \qquad (1)$$

where $t$ is the transport time (also the photoresponse time), $v$ is the velocity of holes, $\mu$ is the mobility of holes, and $E$ is the built-in electrical field defined by $V_{SD}/L$. Therefore we can obtain the relationship of the photoresponse time in BP photodetector with the channel length, in-plane mobility and applied $V_{SD}$. As shown in Supplementary Figs. S9e and S9f, in principle the response speed of the BP photodetector can be as fast as a nano-second (GHz) when using a sub-micron scale channel or applying a larger $V_{SD}$. With the parameters in our virtual BP devices (10 μm channel length, $V_{SD}$ = 0.1 V and hole mobility 500 cm$^2$/V.s), the estimated intrinsic response time from equation (1) will be as small as 0.02 μs.

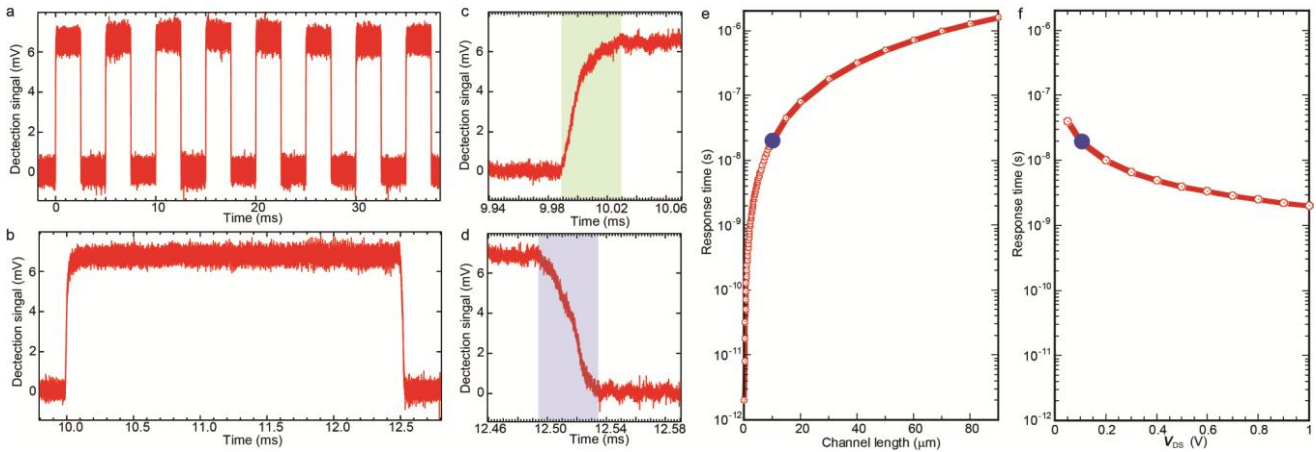

Supplementary Figure S9. Ultrafast time photoresponse in BP linear dichroism photodetector. The photoresponse to eight continuers laser pulses (**a**), single pulse (**b**), laser on process (**c**) and laser off process (**d**). Laser excitation here at 780 nm. **e,** and **f**, Theoretical estimate of the channel length and $V_{DS}$ dependent time response in BP photodetector. Red line indicates theoretical estimate based on equation (1) with experimental parameters. Owing to the high mobility of BP, the BP photodetector can have a very fast photoresponse. Blue dots indicate the estimated photoresponse time for our virtual devices with a channel length of 10 μm and, $V_{SD}$ = 0.1 V and hole mobility 500 cm$^2$/V s.

**Supplementary Section 8: Self-consistent Poisson-Schrödinger calculation for surface band bending in black phosphorus EDLTs**

Self-consistent Poisson-Schrödinger calculations were performed to estimate the band alignment, the wave function extension, and the carrier density distribution for surface band bending at the Gel/BP interface. We solve Poisson's equation within a modified Thomas-Fermi approximation (MTFA, developed by Paasch and Übensee), which considers quantum mechanical effects within the calculations.[4, 5, 6] The density of states, obtained from the *ab initio* band calculation, is modified by including the MTFA factors.[7, 8, 9] The MTFA approximation has been shown to be in excellent agreement with full self-consistent Poisson-Schrödinger solutions for parabolic and non-parabolic conduction band dispersions.[7, 8, 9] Here, we give more details on how we solve Poisson's equation for our particular BP case based on the well-known MTFA model.

First of all, the potential *V*(z) formed at the surface of BP follows the Poisson equation.

$$\frac{d^2V}{dz^2} = -\frac{\rho(z)}{\varepsilon(0)\varepsilon_0}$$

where *z* is the depth from the surface, ρ(z) is the space charge per unit volume in BP, ε(0) is the static dielectric constant of Black Phosphorus (we using the value 8.3 from reference[10]). The space charge is composed of positive and negative charges due to static impurities and mobile electrons and holes, i.e.

$$\rho(z) = e[N_d^+ - n(z) + p(z) - N_a^-]$$

$N_d^+$ is the density of bulk donors while the density of mobile electrons has been described by n(z). $N_a^-$ is the density of bulk acceptors while the density of mobile hole are *ρ*(z). The number of electrons per unit volume in the conduction band of a semiconductor is given by

$$n = \int_0^\infty g_c(E) f_{FD}(E) f(z) dE$$

Here we define the zero of energy at the conduction band minimum (CBM). $g_c$(E) is the density of states for the conduction band. $f_{FD}$ is the Fermi-Dirac function including the potential dependence.

$$f_{FD}(E) = \frac{1}{1+\exp\{\beta[E-E_F+V(z)]\}}$$

where $E_F$ is the Fermi energy, $\beta$ is product of Boltzmann constant and room temperature as $1/k_B T$. $f(z) = 1 - sinc\left[\frac{2z}{L}\left(\frac{E}{k_B T}\right)^{\frac{1}{2}}\right]$ is the MTFA factor to include the effects of quantum-mechanical reflection at this barrier.[7] $L$ is the thermal length as $\hbar/\sqrt{2m^* k_B T}$. Similarly, the density of mobile holes can be expressed as:

$$p = \int_{-\infty}^{E_v} g_v(E)[1-f_{FD}(E)]f(z)dE$$

$g_v(E)$ is the density of states for valence band. $E_v$ is the highest valence band energy. What's more, the potential $V(z)$ must satisfy the boundary condition

$$V(z) \to 0, as\ z \to \infty \quad \text{and} \quad V(0) = V_B$$

where $V_B$ is the band bending at surface.

We use a numerical solution of the Schrödinger equation for the resulting one-electron potential to yield the corresponding eigenfunction for electrons inside the potential well.[7] Then, the carrier concentration profile can be calculated as a sum of the probability of every occupied eigenstate. We use Fourier-series representation to get the numerical solution of Schrodinger equation.

$$[E_c(-i\nabla)+V(z)]\psi = E\psi$$

where $E_c$ is the energy of conduction band electron without any concern of band bending. And $V(z)$ is the one-electron potential. The electron is free along the parallel direction of the surface, and it is confined along the vertical direction. It is therefore appropriate to give wave function in terms of Wannier function $\psi(r_{//},z)$.

$$\psi_{k_{II}}(r_{II}, z) = \frac{1}{\sqrt{A}} \exp(i\vec{k}_{II} \cdot \vec{r}_{II}) \varphi_{k_{II}}(z)$$

Here **r**$_{II}$ represents the parallel component of position vector. The wave function must satisfy the boundary condition that

$$\varphi_{k_{II}}(0) = 0, \varphi_{k_{II}}(l) = 0$$

Thus, $\varphi_{k_{II}}(z)$ can be expanded through Fourier sine series.

$$\varphi_{k_{II}}(z) = \sum_{v=1}^{\infty} \sqrt{\frac{2}{l}} a_v^{k_{II}} \sin(\frac{v\pi}{l} z)$$

Substituting this into the Schrodinger equation and then determining the Fourier coefficients by multiplying sine series and calculating the integration, the result turns out to be matrix representation for a given k$_{II}$.

$$M^{k_{II}} a^{k_{II}} = E^{k_{II}} a^{k_{II}}$$

where the elements of **M** are given by

$$[M]_{vv'} = E_c(k_v)\delta_{vv'} + \frac{2}{l}\int_0^l V(z)\sin(\frac{v\pi}{l}z)\sin(\frac{v'\pi}{l}z)dz$$

where $E_c(k_v)$ can be obtained by equation given before with the $k_v = \sqrt{k_{II}^2 + (v\pi/l)^2}$. And $\delta_{vv'}$ is the Kronecker delta function. The eigenvalues and eigenfunctions of M can therefore be used to determine the confined subband energies and wave functions normal to the surface for a given potential. The subband energies will be discrete in the $k_{II}$ direction but continuous in the $k_\perp$ direction since the confinement is only in the *z* direction.

Therefore we can estimate the band alignment, the wave function extension, and the carrier density distribution for BP surface band bending. The band bending calculations from self-consistent Poisson-Schrödinger solutions for both hole accumulation with upward band bending (supplementary Figs. S10a and S10b) and electron accumulation with downward band bending (Supplementary Figs. S10c and S10d) indicates that carriers are confined in BP within 2 nm from the outmost surface and multiple subbands can be filled. This further indicates that quantum confinement is relatively strong. Note that based on *p*-type BP flakes with initial hole conduction, surface electron accumulation induced by downward band bending can create a vertical n-p junction along the crystal *z* axis, where the built-in electric field perpendicular to the surface will spatially separate photo-generated electrons and holes along *z* direction and reduce their recombination. The exact position of p-n junction will be the location where electron density equal hole density which will be around 10~15 nm from BP surface.

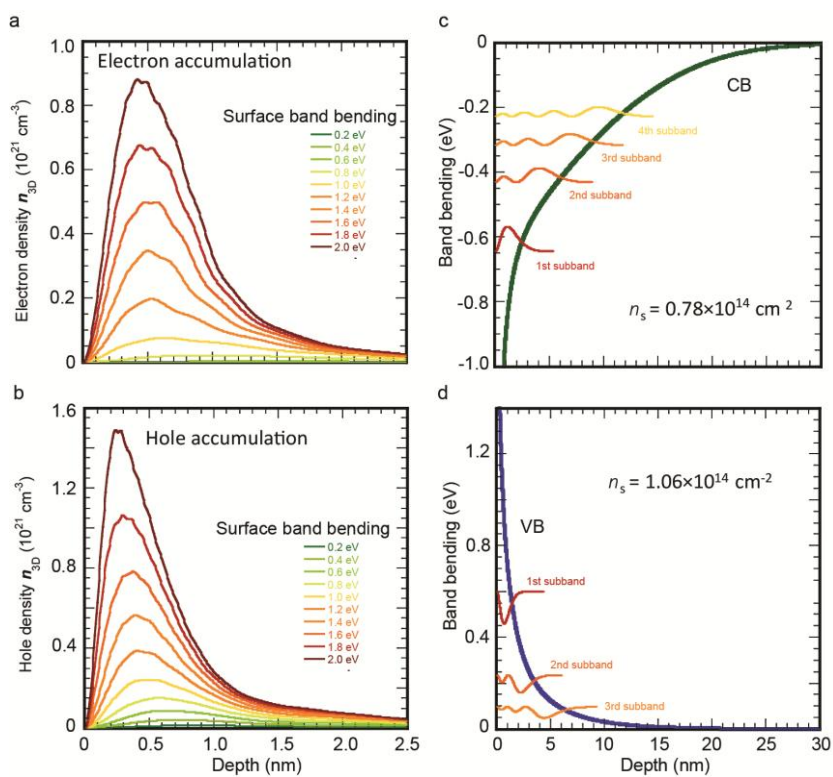

**Supplementary Figure S10.** Self-consistent Poisson-Schrödinger calculations on surface band bending and carrier distribution profile in ionic gel gated BP transistors. **a** and **b,** Carrier distribution profile for electron or hole accumulation. The profiles with different colours represent those scenarios with varied band bending on BP surface assuming there is no interfacial trap states, no impurity states inside the band gap. **c** and **d,** Subband filling with the 2.0 eV downward band bending (**c**) and the 2.0 eV upward band bending (**d**).